\documentclass[nofootinbib,aps,superscriptaddress,preprint]{revtex4}

\usepackage{graphicx}

\def\d{{\rm d}}

\def\D0bar{\overline D{}^0}
\def\K0bar{\overline K{}^0}
\def\DDbar{D^0-\overline D{}^0}

\newcommand{\beq}{\begin{equation}}
\newcommand{\eeq}{\end{equation}}
\newcommand{\beqa}{\begin{eqnarray}}
\newcommand{\eeqa}{\end{eqnarray}}

\begin{document}
\vspace{3.0cm}
\preprint{\vbox {
\hbox{WSU--HEP--0503} \hbox{hep-ph/0506185}}}

\vspace*{2cm}

\title{\boldmath Short Distance Analysis of $D^0-\D0bar$ Mixing}

\author{Eugene Golowich}
\affiliation{Department of Physics, 
	University of Massachusetts\\[-6pt]
	Amherst, MA 01003}

\author{Alexey A.\ Petrov\vspace{8pt}}
\affiliation{Department of Physics and Astronomy\\[-6pt]
	Wayne State University, Detroit, MI 48201\\[-6pt] $\phantom{}$ }

\begin{abstract}
We study the Standard Model short-distance prediction for the mass and 
lifetime differences between the two neutral $D$ meson mass eigenstates.
We find that, despite $\alpha_s/4\pi$ suppression, next-to-leading 
order (NLO) short-distance QCD corrections exceed the 
corresponding leading order (LO) amplitudes.  For the lifetime 
difference, this stems from the lifting of helicity suppression of a 
light-quark intermediate state. 
We find $y_{\rm D}$ is given by $y_{\rm NLO}$ 
to a reasonable approximation but $x_{\rm D}$ is greatly 
affected by destructive interference between $x_{\rm LO}$ and 
$x_{\rm NLO}$.  The net effect is to render $y_{\rm D} \sim x_{\rm D}
\simeq 6 \cdot 10^{-7}$. Our NLO short-distance results, still 
smaller than most long-distance estimates, depend on the same two 
nonperturbative matrix elements of four-quark operators as in 
leading order.  
\end{abstract}

\def\thepage{{}}
\maketitle
\def\thepage{\arabic{page}}

\section{Introduction}
Experimental efforts to detect $\DDbar$ mixing are longstanding and 
remain an active area to this day.~\cite{data1,data2,data3}
The theory of $\DDbar$ mixing is relevant both in lending 
phenomenological guidance to ongoing experimental work and 
in better understanding the workings of the 
Standard Model and of various New Physics 
scenarios~\cite{Petrov:2003un,Burdman:2003rs}.   
In this paper, we present new results  --- the perturbative 
QCD NLO contributions in the framework of the 
$1/m_c$ expansion for 
$\Delta \Gamma_D$ and $\Delta M_D$.  The complex of $D$-meson 
phenomena presents a nontrivial theoretical laboratory for studying 
applicability of heavy quark methods.  One can argue that 
$m_c \gg \Lambda_{\rm QCD}$ justifies the use of heavy quark methods.
However, the scale $\mu \simeq M_D$ lies in the meson resonance 
region, so QCD dynamics is clearly 
present~\cite{Golowich:1998pz}.  As such, there is inherently a degree
of interest in the numerical aspect of our findings.
Our calculation also touches on matters of 
principle, such as the degree of $m_q/m_c$ suppression 
in $\Delta \Gamma_D$ and $\Delta M_D$ at NLO order.

We begin by reviewing the theoretical context of 
$D^0-\D0bar$ mixing.
The mixing arises from $\Delta C=2$ interactions that generate 
off-diagonal terms in the mass matrix for $D^0$ and $\D0bar$ mesons.
The expansion of the off-diagonal terms in the neutral $D$ mass
matrix to second order in perturbation theory is 
\beq\label{M12}
\left (M - \frac{i}{2}\, \Gamma\right)_{12} =
  \frac{1}{2M_D}\, \langle \D0bar | {\cal H}_w^{\Delta C=2} | D^0 \rangle +
  \frac{1}{2M_D}\, \sum_n {\langle \D0bar | {\cal H}_w^{\Delta C=1} | n 
  \rangle\, \langle n | {\cal H}_w^{\Delta C=1} | D^0 \rangle 
  \over M_D-E_n+i\epsilon} \ \ ,
\eeq
where ${\cal H}_w^{\Delta C=2}$ is the effective $\Delta C=2$ hamiltonian 
and ${\cal H}_w^{\Delta C=1}$ is
\beq
{\cal H}_w^{\Delta C=1}=\frac{G_F}{\sqrt{2}} 
\sum_{q,q'}\ V_{cq}^*V_{uq'} 
\left[ C_1(\mu) Q_1 + C_2(\mu) Q_2 \right]\ \ .
\label{hw}
\eeq 
In ${\cal H}_w^{\Delta C=1}$, 
the flavor sum on $q,q'$ extends over the $d,s$ 
quarks,\footnote{In this paper, we work with $m_u = m_d = 0$.} 
the quantities $C_{1,2}(\mu)$ are Wilson coefficients evaluated
at energy scale $\mu$, and $Q_{1,2}$ are the four-quark operators
\beq
Q_1 = \left(\overline{q}_i c_j\right)_{V-A}\,
\left(\overline{u}_i q'_j\right)_{V-A} \qquad 
{\rm and} \qquad 
Q_2 = \left(\overline{q}_i c_i\right)_{V-A}\,
\left(\overline{u}_j q'_j\right)_{V-A}\ \ .
\eeq
The first term in Eq.~(\ref{M12}) represents $\Delta C=2$ 
contributions that are local at 
scale $\mu \sim M_D$, so it contributes to the $M_{12}$ 
(but not to the $\Gamma_{12}$) part of the mixing matrix.   
For example, in the Standard Model this term is generated 
by the contribution of the $b$ quark. It can
also receive a potentially large enhancement from new physics. 
The second term in Eq.~(\ref{M12}) comes from a  
double insertion of $\Delta C=1$ operators in the SM lagrangian, 
and it contributes to both $M_{12}$ and $\Gamma_{12}$.  
It is dominated by SM contributions even 
in the presence of new physics. At scale $\mu \sim M_D$, 
the contributions are from the strange and down 
quarks and these have relatively large CKM factors.
By contrast the $\Delta C=2$ term is expected to give a 
negligible contribution ({\it e.g.} in the SM there is the 
severe CKM suppression 
$|V_{ub} V_{cb}^*|^2/|V_{us} V_{cs}^*|^2 = 
{\cal O}(10^{-6})$).  Thus, we omit it henceforth. 

The off-diagonal mass-matrix terms induce mass eigenstates 
$D_L$ and $D_S$ which are superpositions of the flavor eigenstates $D^0$ and $\D0bar$,
\begin{equation} \label{definition1}
   | D_{L,S} \rangle = p\, | D^0 \rangle \pm q\, | \D0bar \rangle\ \ ,
\end{equation}
where $|p|^2 + |q|^2=1$. In the Standard Model CP violation in $D$ 
mixing is negligible, as is CP violation in $D$ decays both in the 
Standard Model and in most scenarios of new physics.  We therefore 
assume in the rest of this paper that CP is a good symmetry, 
and adopt the phase convention~\cite{Donoghue:1992dd}
\begin{equation} \label{convention}
 {\cal C} {\cal P}  | D^0 \rangle =  - \, | {\bar D}^0 \rangle \ \ ,
\end{equation}
Then we have $p=q$, and $| D_{L,S} 
\rangle$ become the CP eigenstates $| D_{\pm} \rangle$ with  
${\cal C}{\cal P} | D_{\pm} \rangle = \pm | D_{\pm} \rangle$.  
We then define the mass and width differences 
\begin{equation} 
\Delta M_{\rm D} \equiv M_{D_+} - M_{D_-} 
\qquad  {\rm and} \qquad \Delta \Gamma_{\rm D} \equiv \Gamma_{D_+} - 
\Gamma_{D_-} \ \ .
\label{diffs}
\end{equation}
It is, however, customary to work directly with the dimensionless 
quantities, 
\beq
x_{\rm D} \equiv {\Delta M_{\rm D} \over \Gamma_{\rm D}}, \qquad 
y_{\rm D} \equiv {\Delta \Gamma_{\rm D} \over 2\Gamma_{\rm D}},
\eeq
where $\Gamma_{\rm D}$ is the average width of the two 
neutral $D$ meson mass eigenstates.  

The discussion thus far covers relevant background material.  
We conclude this section by 
addressing three particularly important additional points:  
\begin{enumerate}
\item Our calculation adopts an operator product 
expansion (OPE)~\cite{Georgi:1992as,Bigi:2000wn}.  In the limit 
$m_c \gg\Lambda$, where $\Lambda$ is some soft QCD scale, the momentum
flowing through the light degrees of freedom in the intermediate 
state is large.  As such, an OPE is 
implemented by expanding the second term in Eq.~(\ref{M12}) 
in series of matrix elements of local operators. For example, one 
writes for $\Delta \Gamma_{\rm D}$, 
\begin{equation}\label{gammaope}
\Delta \Gamma_{\rm D} = - 2\Gamma_{12} 
= - \frac{1}{M_D}\, {\rm Im}\, \langle \D0bar |
    \,i\! \int\! {\rm d}^4 x\, T \Big\{
    {\cal H}^{\Delta C=1}_w (x)\, {\cal H}^{\Delta C=1}_w(0) \Big\}
    | D^0 \rangle \ \ , 
\end{equation}
and expands the time ordered product in 
Eq.~(\ref{gammaope}) in local operators of 
increasing dimension (higher dimension operators being 
suppressed by powers of $\Lambda/m_c$).
\item We calculate $\Delta \Gamma_{\rm D}$ by making direct use of work 
available in the literature~\cite{Beneke:2003az}, but not 
heretofore applied to $D^0-\D0bar$ mixing.  We then compute the 
mass difference $\Delta M_{\rm D}$ from an unsubtracted 
dispersion relation,\footnote{The tiny $b$-quark contribution, 
neglected here, would contribute to a subtraction constant.}
\beq\label{delmd}
\Delta M_{\rm D} (m_c^2) 
= -\frac{1}{2\pi}\, {\rm P}\! \int_{s_0}^\infty \d s
\  \frac{\Delta\Gamma_{\rm D}(s)}{s-m_c^2}\ \ ,
\eeq
which follows from the analyticity of $\Delta\Gamma_{\rm D}$ 
in the complex $s$-plane with a unitarity branch cut along the 
${\cal R}e ~s$ axis~\cite{Falk:2004wg}. 
\item We expand all our expressions for $x_{|rm D}$ 
and $y_{\rm D}$ in powers of the ratio $z=m_s^2/m_c^2$.  
\end{enumerate}

\section{Analysis}
\begin{figure} [htb]
\includegraphics[height=1.7cm]{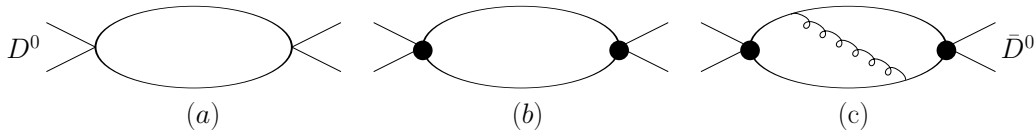}
\caption{$D^0 \to \D0bar$: (a) No-QCD, (b) QCD-corrected vertices, 
(c) an example of NLO correction. \hfill 
\label{fig:fig1}}
\end{figure}

In what follows we compute LO and NLO contributions to 
$y$ and then $x$,
\beq
y_{\rm D}=y_{\rm LO}+y_{\rm NLO} \qquad {\rm and} \qquad 
x_{\rm D}=x_{\rm LO}+x_{\rm NLO} \ \ .
\eeq
We depict in Fig.~\ref{fig:fig1} 
how QCD affects the $D^0$-to-$\D0bar$ mixing amplitude: 
(a) the limit of no QCD 
corrections, (b) the LO component in which QCD dresses the 
interaction vertices, and (c) an example of a NLO correction. 

\vspace{1.0cm}

\vspace{0.5cm}

The leading contribution to $\Delta\Gamma_{\rm D}$ 
in the $1/m_c$ expansion to $\DDbar$ mixing
comes from the dimension-six $|\Delta C|=2$ four-quark operators,
\begin{eqnarray}
Q &=& \bar u_\alpha \gamma_\mu P_L c_\alpha\,
    \bar u_\beta \gamma_\mu P_L c_\beta\,, \qquad
    Q_S = \bar u_\alpha P_L c_\alpha\,
    \bar u_\beta P_L c_\beta\,, \nonumber \\
Q'&=& \bar u_\alpha \gamma_\mu P_L c_\beta\,
    \bar u_\beta \gamma_\mu P_L c_\alpha\,, \qquad
    O_S' = \bar u_\alpha P_L c_\beta\,
    \bar u_\beta P_L c_\alpha \ \ , 
\end{eqnarray}
where $P_L = (1+\gamma_5)/2$. One can use Fierz identities and equations of
motion to eliminate $Q'$ and $Q_S'$ in favor of $Q$ and $Q_S$.  The 
resulting expression for $\Delta\Gamma_{\rm D}$ is then 
\beq\label{delgam1}
\Delta \Gamma_{\rm D} = \frac{G_F^2 m_c^2}{12 \pi M_D}\,
\left[F(z)\, \langle \D0bar | Q(\mu') | D^0 \rangle \, + 
F_S(z)\, \langle \D0bar | Q_S(\mu') | D^0 \rangle \, \right], 
\eeq
Coefficients $F(z)$ and $F_S(z)$ are defined as
\beqa\label{pqcdcoeff}
F(z) &=& \sum_{qq'} \xi_q \xi_{q'} \left[
F_{11}^{qq'}(z) C_1^2(\mu) +  F_{12}^{qq'}(z) C_1(\mu) C_2(\mu) +
F_{22}^{qq'}(z) C_2^2(\mu) \right],
\nonumber \\
F_{ij}^{qq'}(z) &=& F_{ij}^{(0)qq'}(z) 
+ \frac{\alpha_s(\mu)}{4\pi} F_{ij}^{(1)qq'}(z),
\eeqa
and similarly for $F_S(z)$. Here $\xi_{q} \equiv V_{cq}^* V_{uq}$ 
is a CKM factor for the intermediate $s,d$ quarks, the 
$\{ F_{ij}^{(0)qq'}(z) \}$ functions are given in the discussion 
to follow and the $\{ F_{ij}^{(1)qq'}(z) \}$ are considered later in 
our NLO analysis.  As usual, 
the $D^0$-to-$\D0bar$ matrix elements of $Q$ and $Q_S$ 
are parameterized in terms of B-factors, 
\beq
\langle \D0bar | Q | D^0 \rangle \, = \frac{8}{3}f_D^2 M_D^2 B_{\rm D}
\qquad {\rm and} \qquad 
\langle \D0bar | Q_S | D^0 \rangle \, = -\frac{5}{3}f_D^2 M_D^2
{\overline B}_{\rm D}^{(S)} \ \ ,
\label{b-fctr}
\eeq
where ${\overline B}_{\rm D}^{(S)} \equiv B_{\rm D}^{(S)}M_D^2/m_c^2$. 
There are limits on the precision of 
$B_{\rm D}$ and ${\overline B}_{\rm D}^{(S)}$ 
because the calculable short distance component most likely 
gives a negligibly small contribution.  
The most recent result for the quenched lattice calculation of 
$B_{\rm D}$ is reported in Ref.~\cite{Gupta:1996yt}.

\subsection{Leading Order (LO) Contributions}

At leading order in $\alpha_s$, one finds 
for the $s{\bar s}$ intermediate state contributions to 
$F(z)$ and $F_S(z)$,
\beqa
\label{ssfunctions}
\begin{array}{l}
F_{11}^{(0)ss}(z)=3\sqrt{1-4z}~(1-z) \\
F_{12}^{(0)ss}(z)=2\sqrt{1-4z}~(1-z) \\
F_{22}^{(0)ss}(z)=\displaystyle{1 \over 2} (1-4z)^{3/2}
\end{array}
\qquad 
\begin{array}{l}
F_{S11}^{(0)ss}(z)=3\sqrt{1-4z}~(1+2z)\\
F_{S12}^{(0)ss}(z)=2\sqrt{1-4z}~(1+2z)\\
F_{S22}^{(0)ss}(z)=-\sqrt{1-4z}~(1+2z)\ \ ,
\end{array}
\eeqa
and for the $d{\bar s}$ and $s{\bar d}$ contributions, 
\beqa
\label{sdfunctions}
\begin{array}{l}
F_{11}^{(0)ds}(z)=3(1-z)^2\left(1+ \displaystyle{z \over 2}\right) \\
F_{12}^{(0)ds}(z)=2(1-z)^2\left(1+ \displaystyle{z \over 2}\right)\\
F_{22}^{(0)ds}(z)=\displaystyle{1 \over 2} (1-z)^3 
\end{array}
\qquad 
\begin{array}{l}
F_{S11}^{(0)ds}(z)=3(1-z)^2~\left(1+2z\right) \\
F_{S12}^{(0)ds}(z)=2(1-z)^2~\left(1+2z\right) \\
F_{S22}^{(0)ds}(z)=-(1-z)^2~\left(1+2z\right) \ \ .
\end{array}
\eeqa
In addition, we have $F_{ij}^{(0)dd}(z)=F_{ij}^{(0)ss}(0)$. 
Insertion of 
Eqs.~(\ref{pqcdcoeff}),(\ref{ssfunctions}),(\ref{sdfunctions}) 
into Eq.~(\ref{delgam1}) results in the following expression for 
the leading ${\cal O}(z^3)$ contribution, 
\begin{eqnarray}\label{yLO}
y_{\rm LO}^{(z^3)} = {G_F^2 m_c^2 f_D^2 M_D \over 
3 \pi \Gamma_D} ~\xi_s^2~
z^3~ \left( C_2^2 - 2 C_1 C_2 - 3 C_1^2 \right) \left[ 
B_{\rm D} - {5 \over 2} {\overline B}_{\rm D}^{(S)}
\right] \ \ , 
\end{eqnarray}
where $\Gamma_D \simeq 1.6 \cdot 10^{-12}$~GeV is the experimentally 
determined $D^0$ decay rate.  The above expression for 
$y_{\rm LO}^{(z^3)}$ 
agrees in the no-QCD limit of $C_1 = 0$ and $C_2 = 1$ 
with that found in the literature~\cite{dk}.  Since we expect 
$5 {\overline B}_{\rm D}^{(S)}/2 > B_{\rm D}$, it follows that 
$y_{\rm LO} < 0$. 

An expression for $x_{\rm LO}$ is recovered by inserting 
$\Delta\Gamma_{\rm LO}$ into the dispersion relation of 
Eq.~(\ref{delmd}).  One disperses in the variable $m_c^2$ 
so that $z = m_s^2/m_c^2 \to m_s^2/s$.  The functions 
$\{F_{ij}^{(0)}(z)\}$ of 
Eqs.~(\ref{ssfunctions}),(\ref{sdfunctions}) are employed 
above the threshold for each intermediate state.  
Although the dispersion integral diverges separately for each of the 
$s{\bar s}$, $d{\bar d}$, $d{\bar s}$, $s{\bar d}$ intermediate 
states, the flavor-summed expression 
for $\Delta M_{\rm D}$ is rendered finite by GIM cancellations.  
All integrals are first evaluated analytically and the
results are then expanded in powers of $z$.
We find that the leading order in the $z$-expansion 
for $x_{\rm LO}$ occurs at ${\cal O}(z^2)$,
\begin{eqnarray}\label{xLO}
x_{\rm LO}^{(z^2)} &=& 
{G_F^2 m_c^2 f_D^2 M_D \over 3 \pi^2 \Gamma_D} ~\xi_s^2~
z^2~ \bigg[  C_2^2 B_{\rm D} - {5 \over 4} (C_2^2 - 
2 C_1 C_2 - 3 C_1^2) {\overline B}_{\rm D}^{(S)}\bigg] \ \ .
\end{eqnarray}
As with $y_{\rm LO}$, we again regain the standard 
no-QCD result~\cite{{Bigi:2000wn},dk}.  
\begin{table}[t]
\caption{\bf LO Values}
\begin{tabular}{l||c||c|c} 
\colrule 
 &Without QCD  & With QCD (LO in $z$)& With QCD (Exact) \\
\colrule 
$y_{\rm LO}$ &  
$-(2.9 \to 4.8) \cdot 10^{-8}$ & $-(5.6 \to 9.4) \cdot 10^{-8}$ & 
$-(5.7 \to 9.5) \cdot 10^{-8}$ \\
$x_{\rm LO}$ &  
$-(0.53 \to 1.05) \cdot 10^{-6}$ & $-(1.3 \to 2.3) \cdot 10^{-6}$ & 
$-(1.4 \to 2.4) \cdot 10^{-6}$ \\
\colrule 
\end{tabular}
\vskip .05in\noindent
\label{tab:LO}
\end{table}
Terms occurring at next-to-leading order in the $z$-expansion 
are straightforward to determine, and we find 
\begin{eqnarray}
y_{\rm LO}^{(z^4)} &=& 
{G_F^2 m_c^2 f_D^2 M_D \over 3 \pi \Gamma_D} ~\xi_s^2~
z^4~\bigg[ B_{\rm D}  \left( C_2^2 - 4 C_1 C_2 - 6 C_1^2 \right) 
 - {15 \over 4} {\overline B}_{\rm D}^{(S)}
\left( C_2^2 - 2 C_1 C_2 - 3 C_1^2 \right) \bigg]\nonumber \\
 x_{\rm LO}^{(z^3)} &=& 
{G_F^2 m_c^2 f_D^2 M_D \over 3 \pi^2 \Gamma_D} ~\xi_s^2~
z^3~ \bigg[ {1 \over 2} B_{\rm D} \left(C_2^2 + 
2 C_1 C_2 + 3 C_1^2 \right)\\
& & \hspace{1.6cm}  - \ln z ~ \left( B_{\rm D} - {25 \over 12} 
{\overline B}_{\rm D}^{(S)} \right) \left( 
C_2^2 - 2 C_1 C_2 - 3 C_1^2\right) \bigg] 
\ \ .
\label{ntl}
\end{eqnarray}
Notice that at order $x_{\rm LO}^{(z^3)}$, there is now dependence 
also on $\ln z \simeq - 5$. However, these contributions are 
quite small relative to those of Eqs.~(\ref{yLO}),(\ref{xLO}).  

Numerical evaluations for $y_{\rm LO}$ and $x_{\rm LO}$
appear in Table~\ref{tab:LO}.  The initial two 
columns display the leading $z$-dependences first with QCD turned off 
({\it cf} Fig.~\ref{fig:fig1}(a)) and 
then with QCD included via the Wilson coefficients of 
Eq.~(\ref{hw}) ({\it cf} Fig.~\ref{fig:fig1}(b)).    
The final column exhibits the exact LO results.  
The spread of values reflects
uncertainties in input parameters (in particular, we have 
allowed for the range $B_{\rm D}^{(S)}/B_{\rm D} = 0.8 \to 1.2$.  

The collection of LO results in  
Table~\ref{tab:LO} gives rise to several interesting 
questions, but the most obvious one involves the tiny magnitudes.  
The main suppression arises from the presence of 
$z^3 \sim 2 \cdot 10^{-7}$ in $y_{\rm LO}$ and 
$z^2 \sim 4 \cdot 10^{-5}$ in $x_{\rm LO}$, even though 
the expansions for 
$F_{ij}(z)$ and $F_{Sij}(z)$ begin 
at ${\cal O}(1)$.  Such ${\cal O}(1)$ contributions would be 
enormous, but they are in fact cancelled away as are ${\cal O}(z)$ terms.  
As a result, $y_{\rm LO}$ and $x_{\rm LO}$ are rendered tiny.
A numerical by-product of the dependence $y_{\rm LO} \sim {\cal O}(z^3)$ 
and $x_{\rm LO} \sim {\cal O}(z^2)$ is that 
$|x_{\rm LO}| \gg |y_{\rm LO}|$.  
There is of course a corresponding physics explanation.   
In the diagrams of 
Fig.~\ref{fig:fig1}, the $b$-quark contribution is severly 
CKM suppressed, so only the light $s,d$ quarks propagate on internal  
legs.  Since the mixing amplitude will vanish in the $m_d = m_s = 0$ 
limit, the breaking of chiral symmetry and of SU(3) flavor symmetry  
play crucial roles.  Thus, a factor of 
$m_s^2$ comes from an $SU(3)$ violating 
mass insertion on each internal quark line and 
another from an additional mass
insertion on each line to compensate the chirality flip from the first
insertion.  This mechanism of chiral suppression
accounts for the $z^2$ dependence of 
$x_{\rm LO}$.  In addition, $y_{\rm LO}$ requires yet another 
factor of $m_s^2 \propto z$ to lift the helicity suppression 
for the decay of a scalar meson into a massless fermion pair.  


\subsection{Next-to-Leading Order (NLO) Contributions}

Any way of reducing the chiral and helicity supression 
in $x$ and $y$ should lead to an enhancement.  In principle, there 
are both nonperturbative and perturbative ways to achieve this.  

One might associate nonperturbative effects with the presence of 
quark condensates in the QCD vacuum~\cite{Georgi:1992as,Bigi:2000wn}.
These contributions (suppressed by powers of $1/m_c$) lead to 
chirality flip the same way mass insertions do, but have an intrinsic 
scale of $\Lambda\sim 1~\mbox{GeV}\gg m_s$. 
In the realistic case of not-so-large $m_c$, such power suppressions 
are not always sufficient to ensure the smallness of higher order 
contributions. Therefore, Eqs.~(\ref{yLO}) and (\ref{xLO}) 
cannot contribute to leading order in the dual expansion 
in $m_s$ and $1/m_c$ if higher order terms in the $1/m_c$ 
expansion contain lower powers of $z$ than do $x_{\rm LO}$ and 
$y_{\rm LO}$.  It has already been shown that this is indeed the 
case~\cite{Georgi:1992as,Bigi:2000wn}. 

There are also perturbative QCD corrections to $x$ and $y$, but these  
have heretofore not been given serious consideration due to 
the negligibly small LO values for the $\DDbar$ mixing parameters.  
Even taking into account large scale dependence, the LO result 
gives a tiny contribution.  This has stimulated a shift of attention 
towards the computation of the long-distance sector 
with varying degrees of model 
dependence~\cite{Donoghue:hh,Falk:2001hx,Falk:2004wg}. 

But due to their milder dependence on $m_s$, the higher order 
QCD corrections might be able to give relatively large 
contributions~\cite{Petrov:1997ch}.  This occurs, for example, 
in the $c \to s \gamma$ short distance amplitude, which receives a huge 
QCD correction~\cite{ghmw}.  In this paper, we consider a 
specific means by which the helicity supression 
in $y$ can be lifted --- a perturbative gluon correction ({\it e.g.} as in
Fig.~\ref{fig:fig1}(c)). Having a perturbative gluon traversing the 
graph for the correlation function is the same as a well-known 
effect of lifting of helicity supression which follows from having three particles 
in the intermediate state instead of 
two~\footnote{This mechanism leads to the prediction that the rate for
the weak radiative decay $B\to \gamma e\nu$ is much larger than 
the rate of weak leptonic
decay $B\to e\nu$}. The addition of the `intermediate-state' 
gluon can lift one power of $z$, which characterizes the 
helicity suppression in $y$. Also, the 
relative lightness of $m_c$ 
implies that higher order perturbative 
QCD corrections are suppressed 
by a relatively large factor of $\alpha_s(m_c) \sim 0.4$. 
It is therefore expected that
the NLO corrections to $y$ should dominate the LO result. 
Moreover, the existence of a 
dispersion relation implies that $x$ might well be 
enhanced at NLO.

In order to systematically include the effects 
of intermediate-state gluons, a complete 
calculation of NLO corrections to $\DDbar$ mixing is needed. 
The NLO corrections to 
lifetime difference $y$ can be readily computed. All the 
relevant NLO contributions to
$F(z)$ and $F_S(z)$ for two massive quarks and 
one massive, one massless quark 
can be found by adopting the formulas 
in Refs.~\cite{Beneke:2003az} (which considered the case of 
$B_s-{\bar B}_s$ mixing) to computing $F_{ij}^{(1)qq'}$ of 
Eq.~(\ref{pqcdcoeff}). That calculation has been 
performed in the NDR-scheme 
(dimensional regularization with anti-commuting 
$\gamma_5$ and $\overline{MS}$
subtraction). We shall not present explicit formulas 
for the $\{ F_{ij}^{(1)qq'}(z) \}$ and 
$\{ F_{S,ij}^{(1)qq'}(z) \}$ functions 
as they are rather cumbersome. 

Scale dependent quantities used in our numerical work and 
evaluated at $\mu = 1.3$~GeV were:
\beq
m_c = 1.3~{\rm GeV}~, \quad B_{\rm D} = 0.82~, \quad 
C_1 = -0.411~, \quad C_2 = 1.208~, \quad \alpha_s = 0.406 \ \ .
\label{values}
\eeq
The value for $B_{\rm D}$ at scale $\mu = 1.3$~GeV is obtained by 
referring the lattice determination at $\mu = 2$~GeV and employing 
the scale invariant quantity ${\hat B}_{\rm D}$, 
\beq
{\hat B}_{\rm D} = B_{\rm D}(\mu_0) [ \alpha_s(\mu_0)]^{-6/25} 
\left[ 1 + {\alpha_s (\mu_0) \over 4 \pi} ~J_4 \right] \ \ , 
\label{bd}
\eeq
with $J_4 \simeq 1.792$.  Also we allow for a range of the ratio 
${\overline B}_{\rm D}^{(S)}/B_{\rm D}$. 

Using the $\{ F_{ij}^{(1)qq'}(z) \}$ and 
$\{ F_{S,ij}^{(1)qq'}(z) \}$ functions, 
we have calculated $y_{\rm NLO}$ exactly and also 
have expressed it in terms of a power series in $z$.  
The leading term is ${\cal O}(z^2)$, 
\beqa
y_{\rm NLO}^{(2)} &=&
{ G_F^2 m_c^2 f_D^2 M_D \over 3 \pi \Gamma_D} ~\xi_s^2~
{\alpha_s \over 4 \pi}~ z^2 
\Bigg( B_{\rm D} \left[ - \left({77\over 6} - {8\pi^2 \over 9}\right)~ C_2^2
+ 14 ~C_1 C_2 + 8 ~C_1^2 \right] \nonumber \\
& & \hspace{0.9cm} 
- {5 \over 2} {\overline B}_{\rm D}^{(S)} ~\left[ 
\left({8 \pi^2 \over 9} - {25 \over 3} \right) ~C_2^2 
+ 20 ~C_1 C_2 + 32~ C_1^2\right] \Bigg) \ \ ,
\label{yNLO2}
\eeqa
and the corresponding ${\cal O}(z^3)$ contribution is 
\beqa
& & y_{\rm NLO}^{(3)} =
{ 2G_F^2 m_c^2 f_D^2 M_D \over 3 \pi \Gamma_D} ~\xi_s^2~
{\alpha_s \over 4 \pi}~ z^3 \nonumber \\
& & \times \Bigg( B_{\rm D} \left[ (15 + 7 \ln z)~ C_2^2
- \left({77 \over 9} + {103 \over 3} \ln z\right)~ C_1 C_2 - (
18 + 58 \ln z)~ C_1^2 \right] \nonumber \\
& & 
- {5 \over 2} {\overline B}_{\rm D}^{(S)} ~\left[ \left({28 \over 3} 
+ 6 \ln z \right) ~ C_2^2 
+ \left({49\over 9} - {118 \over 3}\ln z\right)~ C_1 C_2 - 
\left({31 \over 3} + 58 \ln z \right)~ C_1^2\right] \Bigg) \ \ .
\eeqa
The numerical results, displayed in Table II, reveal that 
$y_{\rm NLO}$ is almost an order of magnitude larger than 
$y_{\rm LO}$ and that the subleading term $y_{\rm NLO}^{(3)}$ 
is smaller than $y_{\rm NLO}^{(2)}$ but not at all negligible.

\begin{table}[t]
\caption{\bf NLO Values}
\begin{tabular}{c|c|c||c||c} 
\colrule 
$y_{\rm NLO}^{(z^2)}$   & $y_{\rm NLO}^{(z^3)}$ & $y_{\rm NLO}$ 
& $x_{\rm NLO}$ & $y_{\rm NLO}^{\rm (PENG)}$ \\
\colrule 
$(2.2 \to 6.3)\cdot 10^{-7}$  &  $(1.7 \to 2.8)\cdot 10^{-7}$
& $(3.9 \to 9.1)\cdot 10^{-7}$ &  $(1.7 \to 3.0)\cdot 10^{-6}$
&  $(0.6 \to 0.8)\cdot 10^{-9}$ \\
\colrule 
\end{tabular}
\vskip .05in\noindent
\label{tab:NLO}
\end{table}

The corresponding expression for 
$x_{\rm NLO}$ has, as before, been obtained by means of a dispersion
relation.  We evaluated the dispersion integral numerically to 
obtain the value presented in Table II.  As regards an analytical 
expression for $x_{\rm NLO}$, the intent was again to 
by first exactly perform the dispersion integrals 
and then expand each contribution in a $z$ power series. 
It turned out possible to do this for the $d{\bar d}$, $d{\bar s}$ 
and $s{\bar d}$ intermediate states, but not for $s{\bar s}$.  
It is, however, nonetheless useful to have 
an approximate analytic representation for $x_{\rm NLO}$.
By exploring a variety of approximation techniques, 
we found the expected ${\cal O}(z^2, z^2 \ln z)$ leading 
behavior for $x_{\rm NLO}$ but encountered scatter 
in the ${\cal O}(z^2)$ coefficients, although less so for the 
${\cal O}(z^2 \ln z)$ coefficients.  Upon accepting the latter 
and fitting the ${\cal O}(z^2)$ coefficients to the numerical 
evaluations of individual dispersion integrals, 
we arrived at the `effective' formula:
\begin{eqnarray}
& & x_{\rm NLO} \simeq 
-{G_F^2 m_c^2 f_D^2 M_D \over 3 \pi^2 \Gamma_D} ~\xi_s^2~ 
{\alpha_s \over 4 \pi} ~ z^2  \nonumber \\
& & \times \bigg( 
B_{\rm D} \left[ (11.3 - 4.1 \ln z)~ C_2^2 + 
(49.2 + 15.8 \ln z) ~C_1 C_2 + (37.9 + 10.7 \ln z)~ C_1^2\right] 
\label{xNLO} \\
& & - {5 \over 8} {\overline B}_{\rm D}^{S} 
\left[ (37.9 + 2.2 \ln z)~C_2^2 + (- 33. + 81.8 \ln z )~C_1 C_2 
+ (32.0 + 125.3 \ln z)~ C_1^2 \right]
\bigg) \ . \nonumber 
\end{eqnarray}
This relation, although approximate, is nonetheless useful in 
understanding the magnitude of the various contributions to 
$x_{\rm NLO}$.  

Since the NLO results found for the box contributions are 
larger than their LO counterparts, we consider here for the 
sake of completeness the 
NLO penguin contribution $y_{\rm NLO}^{\rm (P)}$ to the 
width difference.  We have 
\begin{eqnarray}\label{PNLO}
& & y_{\rm NLO}^{\rm (P)} = 
-{4G_F^2 m_c^2 f_D^2 M_D \over 9 \pi \Gamma_D} ~\xi_s^2~ 
{\alpha_s \over 4 \pi}
~z^3 ~C_2^2 ~\left( B_{\rm D} + 5 {\overline B}_{\rm D}^{S} 
\right) + \dots \ \ .
\end{eqnarray}
The result shown in Table~\ref{tab:NLO} clearly shows 
the penguin amplitude for $y_{\rm NLO}^{\rm (P)}$ is 
negligible compared to the box contribution.  The 
mass splitting $x_{\rm NLO}^{\rm (P)}$ is likewise 
${\cal O}(z^3)$ and hence negligible.  

\section{Concluding Comments}

We have calculated LO and NLO contributions to the leading 
dimension-six component in the OPE for $D^0-\D0bar$ mixing.
Numerical results appear in Table~\ref{tab:LO} for LO 
and Table~\ref{tab:NLO} for NLO.  
As a partial check of our analysis, we found our results 
(in cases of overlap) to agree with work carried out previously.  
Our formulae for $x$ and $y$ involve not simply expansions 
in $1/m_c$, but rather combined expansions in $m_s$ ($m_d$ is negligible), 
$\alpha_s$, and $1/m_c$.  As a technical aside, we performed the 
calculations at scale $m_c \simeq 1.3$~GeV.

The two most noteworthy numerical features found for $x$ and $y$ are:
\begin{enumerate}
\item They are small at LO and even at NLO.  
This is because $z \equiv m_s^2/m_c^2$ 
is small and the leading dependence on $z$ is found to be 
\beq
y_{\rm LO} \sim z^3\qquad x_{\rm LO} \sim z^2\qquad y_{\rm NLO} \sim
z^2\qquad x_{\rm NLO} \sim z^2 \ \ .
\label{order}
\eeq
Although contributions from individual intermediate 
states are {\it not} small, CKM factors cancel away 
the ${\cal O}(1)$ and ${\cal O}(z)$ components.   
\item The NLO terms are larger than the LO terms.  This 
requires somewhat more explanation, especially since NLO 
amplitudes contain the small perturbative QCD factor 
$\alpha_s/4\pi$.  As regards the dimensionless width 
difference $y_{\rm D}$, 
the ratio of leading terms in the $z$ expansion is 
\beq
{y_{\rm NLO}^{(z^2)} \over y_{\rm LO}^{(z^3)}} = 
{\alpha_s \over 4 \pi} \times {1 \over z} \times {W_y^{\rm (NLO)} \over
W_y^{\rm (LO)}} 
\simeq 0.03 \times 169 \times (-0.73) \simeq -4 \ \ .
\label{ratio-y}
\eeq
In the above $W_y^{\rm (NLO)}/W_y^{\rm (LO)}$ is the ratio of terms 
containing the Wilson coefficients in Eqs.~(\ref{yLO}),(\ref{yNLO2}) 
and for definiteness we have considered the case $B_{\rm D}^{(S)} 
= 0.8 B_{\rm D}$.  Eq.~(\ref{ratio-y}) shows that 
$|y_{\rm NLO}^{(z^2)}|$ exceeds $|y_{\rm LO}^{(z^3)}|$ 
because the extra factor of $z$ overwhelms the $\alpha_s/4\pi$ 
suppression.  We have already discussed 
the physics of this -- the helicity suppression mechanism which
affects any LO $q{\bar q}$ intermediate state is removed via the 
presence of a virtual gluon in the NLO $q{\bar q}G$ 
intermediate state.  Also, the difference in sign between 
$y_{\rm NLO}^{(z^2)}$ and $y_{\rm LO}^{(z^3)}$ arises from 
the factor $W_y^{\rm (NLO)}/W_y^{\rm (LO)}$.

Since, to leading order in $z$, $x_{\rm NLO}$ and $x_{\rm LO}$ 
both behave as $z^2$, something else must account for the result 
$|x_{\rm NLO}| > |x_{\rm LO}|$.  From Eq.~(\ref{xLO}) and the 
approximate formula Eq.~(\ref{xNLO}), we have 
\beq
{x_{\rm NLO} \over x_{\rm LO}^{(z^2)}} \simeq 
- {\alpha_s \over 4 \pi} \times {W_x^{\rm (NLO)} \over
W_x^{\rm (LO)}} 
\simeq - 0.03 \times (41.4) \simeq - 1.3 \ \ ,
\label{ratio-x}
\eeq
where $W_x^{\rm (NLO)}$ and $W_x^{\rm (LO)}$ are again the 
contributions from the Wilson constants and their coefficients. 
In this case, the suppression in $\alpha_s/4\pi$ 
is overcome by the large size of $W_x^{\rm (NLO)}/
W_x^{\rm (LO)}$.  In particular, the largest contributor to 
$W_x^{\rm (NLO)}$ is from the ${\overline B}_{\rm D}^{S}$ 
term in Eq.~(\ref{xNLO}), roughly equally between log and 
non-log terms.    
\item We conclude that, citing just central values, the net 
effect of the short distance contributions is 
\beq
y_{\rm D} = y_{\rm LO} + y_{\rm NLO} \simeq 6 \cdot 10^{-7} \ , 
\qquad x_{\rm D} = x_{\rm LO} + x_{\rm NLO} \simeq 6 \cdot 10^{-7} \ .
\label{net}
\eeq
In brief, $y_{\rm D}$ is given by $y_{\rm NLO}$ 
to a reasonable approximation but $x_{\rm D}$ is greatly 
affected by destructive interference between $x_{\rm LO}$ and 
$x_{\rm NLO}$.  The net effect is to render $y_{\rm D}$ and 
$x_{\rm D}$ of similar magnitudes, at least at this order of analysis.
\end{enumerate}

$\DDbar$ mixing thus provides a concrete example 
of a well-defined observable for which 
NLO perturbative QCD corrections dominate the LO result.
Will it follow that the NNLO contributions are larger still?  
Of course, one cannot know without doing the calculation.
We feel, however, it may not necessarily be the case, at least 
as regards the width difference $y$. 
The three-particle $q{\bar q}G$ intermediate states were 
able to lift the helicity suppression experienced by 
$q{\bar q}$ intermediate states.  In passing to the NNLO 
sector, there is no analogous suppression factor to be 
lifted.  Of course, there is always the possibility that large 
numerical coefficients can overturn the $\alpha_s/4\pi$ counting.  

The question remains -- just how large is $D^0-\D0bar$ mixing?
Evidently, it is still not possible to provide a definitive 
theoretical answer and experiment will presumably decide the 
issue.  On a relative basis, our `short-distance' 
numerical results are smaller than most `long-distance' 
estimates (although the model dependence and 
uncertainty present in even modern and improved versions of the latter 
is less significant here).  Experimentalists might find it useful to
interpret our numerical NLO values as lower bounds to $y_{\rm D}$  
and $x_{\rm D}$.

We conclude by considering our analysis in the context of operator product 
expansions. As we have seen above, the prediction of $x$ and $y$ 
is a result of expanding the correlation function 
Eq.~(\ref{gammaope}) in terms of {\it three} `small' quantities, 
$z$, $\Lambda/m_c$, and $\alpha_s$. Since the
first quantity is significantly smaller than the other two, the structure
of the series is rather different from other (usual) applications of 
the OPE, {\it e.g.} $B^0-\overline{B}^0$ mixing or 
$b$-hadron lifetimes~\cite{Gabbiani:2004tp}. 

Working with this combined expansion, we computed the leading contribution 
originating from matrix elements of dimension-six operators. These matrix
elements are commonly parameterized in terms of the two nonperturbative
parameters, $B_D$ and $\overline{B}_D$. The applications of techniques of 
lattice and QCD sum rule evaluations of these operators 
can hopefully further improve the precision of our prediction.

At higher orders in this expansion one would need to take into account
${\cal O}(z^{3/2})$ corrections (multiplied by about a dozen matrix
elements of dimension-nine operators) and ${\cal O}(z)$ corrections 
(with more than twenty matrix elements of dimension-twelve operators). 
This would introduce a veritable multitude of unknown parameters whose 
matrix elements cannot be computed at this time. Simple dimensional 
analysis~\cite{Bigi:2000wn} suggests magnitudes 
$x_D \sim y_D \sim 10^{-3}$, but
order-of-magnitude cancellations or enhancements are possible.
However, any effect of higher orders in $1/m_c$ or $\alpha_s(m_c)$ 
which could render the result to be proportional to $z^n$ in the lowest possible
power $n=1$~\cite{Falk:2001hx} would presumably 
produce a dominant contribution to the prediction of $x$ and $y$.

\acknowledgments

The work of E.G. has been supported in part by the U.S.\ National Science 
Foundation under Grant PHY--0244801.
A.P.~was supported in part by the U.S.\ National Science Foundation under
Grant PHY--0244853, and by the U.S.\ Department of Energy under Contract
DE-FG02-96ER41005.  We thank Sandip Pakvasa for conversations at the 
initial stages of this project.


\end{document}